# Why Watching Movie Tweets Won't Tell the Whole Story?

Felix Ming Fai Wong
EE, Princeton University
mwthree@princeton.edu

Soumya Sen
EE, Princeton University
soumyas@princeton.edu

Mung Chiang
EE, Princeton University
chiangm@princeton.edu

## ABSTRACT

Data from Online Social Networks (OSNs) are providing analysts with an unprecedented access to public opinion on elections, news, movies etc. However, caution must be taken to determine whether and how much of the opinion extracted from OSN user data is indeed reflective of the opinion of the larger online population. In this work we study this issue in the context of movie reviews on Twitter and compare the opinion of Twitter users with that of the online population of IMDb and Rotten Tomatoes. We introduce new metrics to show that the Twitter users can be characteristically different from general users, both in their rating and their relative preference for Oscar-nominated and non-nominated movies. Additionally, we investigate whether such data can truly predict a movie's box-office success.

## Categories and Subject Descriptors

H.1.2 [**Information Systems**]: User/Machine Systems—*Human factors*

## 1. INTRODUCTION

Online social networks (OSN) provide a rich repository of public opinion that are being used to analyze trends and predict outcomes. But such practices have been criticized as it is unclear whether polling based on OSNs data can be extrapolated to the general population [2]. Motivated by this need to evaluate the "representativeness" of OSN data, we report on a study of movie reviews in Twitter and compare them with other online rating sites (e.g., IMDb and Rotten Tomatoes) by introducing new metrics on inferrability ($\mathcal{I}$), positiveness ($\mathcal{P}$), bias ($\mathcal{B}$), and hype-approval ($\mathcal{H}$).

Although the context of this work is movie reviews in Twitter, its scope extends to other product categories and social networks. Twitter is our choice for this study because marketers consider brand interaction and information dissemination as a major aspect of Twitter. The focus on movies in this paper is also driven by two key factors:

(a) *Right in the Level of Interest:* Movies tend to generate a high interest among Twitter users as well as in other online user population (e.g., IMDb).

(b) *Right in Timing:* We collected Twitter data during Academy Award season (Oscars) to obtain a unique dataset to analyze characteristic differences between Twitter and IMDb or Rotten Tomatoes users in their reviews of Oscar-nominated versus non-nominated movies.

We collected data from Twitter between February-March 2012 and manually labeled 10K tweets as training data for a set of classifiers based on SVM. We focus on the following questions to investigate whether Twitter data is sufficiently representative and indicative of future outcomes:

• Are there more positive or negative reviews about movies on Twitter?

• Do users tweet before or after watching a movie?

• How does the proportion of positive to negative reviews on Twitter compare to those from other movie rating sites (e.g., Rotten Tomatoes, IMDb)?

• Do the opinions of Twitter users about the Oscar-nominated and non-nominated movies differ quantitatively from these other rating sites?

• Does greater hype and positive reviews on Twitter directly translate to a higher rating for the movie in other rating sites?

• How well do reviews on Twitter and other online rating sites correspond to box-office gains or losses?

The paper is organized as follows: Section 2 reviews related work. Section 3 discusses the data collection and classification techniques used. The results are reported in 4, followed by conclusions in Section 5.

## 2. RELATED WORK

This work complements earlier works in three related topics: (a) OSNs as a medium of information dissemination, (b) sentiments analysis, and (c) Twitter's role in predicting movies box-office.

**Network Influence.** Several works have reported on how OSN users promote viral information dissemination [10] and create powerful electronic "word-of-mouth" (WoM) effects [7] through tweets. [9, 12] study these tweets to identify social interaction patterns, user behavior, and network growth. Instead, we focus on the sentiment expressed in these tweets on popular new movies and their ratings.

**Sentiment Analysis & Box-office Forecasting.** Researchers have mined Twitter dataset to analyze public reaction to various events, from election debate performance [5] to movie box-office predictions on the release day [1]. In contrast, we improve on the training and classification techniques, and specifically focus on developing new metrics to ascertain whether opinions of Twitter users are sufficiently representative of the general online population of sites like IMDb and Rotten Tomatoes. Additionally, we also revisit the issue of how well factors like hype and satisfaction reported in the user tweets can be translated to online ratings and eventual box-office sales.

## 3. METHODOLOGY

### 3.1 Data Collection

From February 2 to March 12, we collected a set of 12 million tweets (world-wide) using the Twitter Streaming API[1]. The tweets were collected by tracking keywords in the titles of 34 movies, which were either recently released (January earliest) or nominated for the Academy Awards 2012 ("Oscars"). The details are listed in Table 1.

There were two limitations with the API. Firstly, the server imposes a rate limit and discards tweets when the limit is reached, but fortunately, the number of dropped tweets accounts for only less than 0.04% of all tweets, and rate limiting was observed only during the night of the Oscars award ceremony. The second problem is the API does not support exact keyword phrase matching. As a result we received many spurious tweets with keywords in the wrong order, e.g., tracking "the grey" returns the tweet "a cat in the grey box". To account for variations in spacing and punctuation, we used regular expressions to filter for movie titles, and after that we obtained a dataset of 1.77 million tweets.

On March 12, we also collected data from IMDb and Rotten Tomatoes for box office figures and the proportion of positive user reviews per movie.

**Definition of a Positive Review.** In contrast to our classification of user tweets with a movie review as being positive or negative, the users of the above two sites attach a numerical rating to each movie. Therefore, we use Rotten Tomatoes' definition of a "positive" rating as a binary classifier to convert the movie scores for comparison with the data from Twitter. Rotten Tomatoes defines a review being positive when its numerical rating is 3.5/5 or above, and the site also provides the proportion of positive user reviews. Then for IMDb, we can do a simple scaling to give a compatible definition of a positive review as one with a rating of 7/10 or above. The proportion of positive user reviews in IMDb is calculated over the per-movie rating distributions provided in their website.

[1] https://dev.twitter.com/docs/streaming-api

| ID | Movie Title[2] | Category[3] |
|---|---|---|
| 1 | The Grey | I |
| 2 | Underworld: Awakening | I |
| 3 | Red Tails | I |
| 4 | Man on a Ledge | I |
| 5 | **Extremely Loud & Incredibly Close** | I |
| 6 | Contraband | I |
| 7 | **The Descendants** | I |
| 8 | Haywire | I |
| 9 | The Woman in Black | I |
| 10 | Chronicle | I |
| 11 | Big Miracle | I |
| 12 | The Innkeepers | I |
| 13 | Kill List | I |
| 14 | W.E. | I |
| 15 | The Iron Lady | II |
| 16 | **The Artist** | II |
| 17 | **The Help** | II |
| 18 | **Hugo** | II |
| 19 | **Midnight in Paris** | II |
| 20 | **Moneyball** | II |
| 21 | **The Tree of Life** | II |
| 22 | **War Horse** | II |
| 23 | **A Cat in Paris** | II |
| 24 | **Chico & Rita** | II |
| 25 | **Kung Fu Panda 2** | II |
| 26 | **Puss in Boots** | II |
| 27 | **Rango** | II |
| 28 | The Vow | III |
| 29 | Safe House | III |
| 30 | Journey 2: The Mysterious Island | III |
| 31 | Star Wars I: The Phantom Menace | III |
| 32 | Ghost Rider: Spirit of Vengeance | III |
| 33 | This Means War | III |
| 34 | The Secret World of Arrietty | III |

**Table 1: List of movies tracked** (Ref. footnotes 2,3)

### 3.2 Tweet Training & Classification

We classify tweets by relevance, sentiment and temporal context as defined in Table 2.

We highlight several design challenges before describing the implementation. Some of the movies we tracked have terse titles with common words (*The Help*, *The Grey*), and as a result many tweets are irrelevant even though they contain the titles, e.g., "thanks for the help" is a valid tweet. Another difficulty is the large number of non-English tweets. Presently we treat them as irrelevant, but we intend to include them in future work. Lastly, both movie reviews and online social me-dia have their specific vocabulary, e.g., "sick" being used to describe a movie in a positive sense, and this can make lexicon-based approaches common in the sentiment analysis literature [13] unsuitable.

[2] Bold indicates the movie was nominated for the Academy Awards for Best Picture or Best Animated Feature Film.
[3] Trending category: (I) trending as of Feb 2; (II) trending as of Feb 7 after Oscars nomination; (III) trending as of Feb 15 after Valentine's Day.

| Class | Definition | Example |
|---|---|---|
| **Relevance** | | |
| Irrelevant (**I**) | Non-English (possibly relevant), or irrelevant from the context | "thanks for the help" |
| Relevant (**R**) | Otherwise | "watched The Help" |
| **Sentiment** | | |
| Negative (**N**) | Contains *any* negative comment | "liked the movie, but don't like how it ended" |
| Positive (**P**) | *Unanimously* and *unambiguously* positive | "the movie was awesome!" |
| Mention (**M**) | Otherwise | "the movie was about wolves" |
| **Temporal Context** | | |
| After (**A**) | After watching as inferred from context | "had a good time watching the movie" |
| Before (**B**) | Before watching movie | "can't wait to see the movie!" |
| Current (**C**) | Tweeted when person was already inside the cinema | "at cinema about to watch the movie" |
| Don't know (**D**) | Otherwise | "have you seen the movie?" |

Table 2: Definition of tweet classes.

To filter irrelevant and non-English tweets while accounting for Twitter-movie-specific language, we decided to take a supervised machine learning approach for tweet classification, i.e., learn by example. In particular, for each of three meta-classes we train one classifier based on Support Vector Machines (SVMs).

**Preprocessing.** For each line of text, we remove usernames, and convert (1) question and exclamation marks, (2) emoticons, (3) URLs (hinting the tweet carries no sentiment) and (4) isolated @ signs (often used to indicate presence at a physical location) to their corresponding meta-words for the next step.

**Feature Vector Conversion.** Using the MALLET toolkit [11], a line of preprocessed text is converted to a binary feature vector, such that an element is 1 if and only if the corresponding word or meta-word from the previous step exists in the text. Stopwords, e.g., "the", are not removed as opposed to usual practice because some of them are common in movie titles.

**Training and Classification.** We randomly sampled 10,975 tweets and labeled them according to the classification in Table 2. Then we implemented and trained the three classifiers with $SVM^{light}$ [8] and its multiclass variant [4]. Finally, we use them to classify the remaining 1.7 million unlabeled tweets.

We compare our classifiers with three baseline classifiers: a *random* one that assigns a class uniformly at random, a *majority* one that assigns to each tweet the most represented class in the training set, and the Naive Bayes classifier implemented in MALLET. Evaluation is done using 10-fold cross validation using the accuracy rate, i.e., the ratio of the number of correctly classified tweets to the total number. The results in Table 3 indicate that our SVM-based classifiers outperform the baselines by a significant margin.

Although evaluation results can not be compared across papers because different datasets, we note that the 98% sentiment classifier in [1] was tested on tweets that have distinctive movie titles (avoiding the movie *2012*) and are relatively easy to classify (unanimously voted by three MTurk workers to be in the same class).

|  | Relevance | Sentiment | Timing |
|---|---|---|---|
| Random | 0.5 | 0.33 | 0.25 |
| Majority | 0.52 | 0.55 | 0.34 |
| Naive Bayes | 0.89 | 0.74 | 0.73 |
| SVM | 0.93 | 0.78 | 0.78 |

Table 3: Comparison of tweet classifiers

|  | N | P | M |
|---|---|---|---|
| A | 0.045 | 0.13 | 0.12 |
| B | 0.011 | 0.17 | 0.17 |
| C | 0.0019 | 0.019 | 0.090 |
| D | 0.0097 | 0.034 | 0.20 |

Table 4: Fraction of tweets in joint-classes.

## 4. DATA ANALYSIS

In this section, we analyze the Twitter user data to characterize whether they are sufficiently representative of the general online population. In particular, we compare the proportion of positive and negative tweets to the ratings about movies from other online rating sites, e.g., Rotten Tomatoes and IMDb. We introduce metrics to quantitatively characteristic how different Twitter user reviews were from these other sites, and analyze the relationship to box-office sales.

### 4.1 Movie Review Statistics

Out of the 1.77M tweets, 51% of them are classified as irrelevant, and we focus on the remaining 49% in the remaining of this paper. We use the tweet classification of Table 2 to infer the temporal context of user tweets. Figure 1(a) shows that a large proportion of tweets about popular movies are made before watching the movie, e.g., *The Women In Black* (9), *Chronicle* (10), *The Vow* (28), etc. Moreover, as shown in Figure 1(b), most tweets are helpful in publicizing the movies (i.e., WoM) as they often mention screening venues and contain positive opinions. Table 4 shows the joint tweet distribution by sentiment and temporal context. If a person tweets before or after watching, the tweet is likely positive. Tweets sent current to watching are mostly neutral "check-in's" using location-based social networking services.

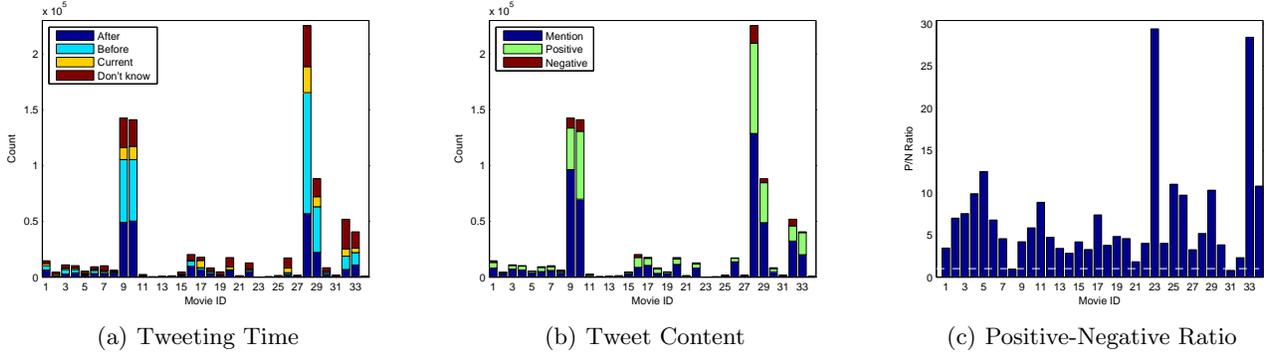

(a) Tweeting Time  (b) Tweet Content  (c) Positive-Negative Ratio

Figure 1: Count of tweets (a) by temporal context, (b) by sentiment, and (c) $P/N$ ratio of movies

Analyzing the impact of these positive and negative online reviews is an important topic for both networking and marketing communities. Product ratings on sites like Amazon typically have a large number of very high and very low scores, which create $J$-shaped histograms over the rating scale [6]. This is attributed to the "brag-and-moan" phenomenon among reviewers. But researchers have also suggested that due to risk-averseness among consumers, negative reviews tend to have a higher impact than positive reviews. However, the impact of these negative reviews can be greatly diminished if they are vastly outnumbered by positive reviews. Hence, it is important to examine whether positive reviews dominate in proportion to negative reviews on OSNs like Twitter.

Figure 1(c) shows that the number of positive reviews on Twitter indeed exceeds the number of negative reviews by a large margin for almost all the movies tracked[4]. Such a large positive bias may be due to the psychology of sharing positive and helpful image among followers. This observation holds some promising implication in developing general marketing strategies for sellers and distributors. For example, instead of focusing on reducing the negative reviews from a few dissatisfied customers, it may be better to focus on enhancing the already high proportion of positive reviews on OSNs and use virality effects to influence consumers.

### 4.2 Movie Preferences of Twitter Users

In this section we compare the proportions of positive to negative user reviews in Twitter, IMDb and Rotten Tomatoes. In Twitter, a positive review is a tweet in class $AP$, and a negative review is a tweet in class $AN$. Thus the proportion of positive to negative reviews in Twitter is the ratio $\frac{AP}{AP+AN}$. Our stringent definition of a tweet being positive, i.e., not containing *any* negative comment, makes the ratio an underestimate of the actual proportion, and as we will see, can only strengthen our results. We also contrast our definition to existing work on sentiment analysis, which can only identify the ratio $\frac{P}{P+N}$ and is likely to overestimate the proportion of positive reviews because of the dominance of positive tweets.

**Qualitative Results.** Figures 2(a) to 2(d) show the scatter plots of the proportions of positive reviews across Twitter, IMDb and Rotten Tomatoes. The dotted lines in the plots makes an angle $\pi/4$ (in radians) with the x-axis and indicate the location the proportion being the same in Twitter or IMDb/Rotten Tomatoes, i.e., if a datapoint is above the line, then users in IMDb/Rotten Tomatoes are more positive than those in Twitter on a certain movie, and vice versa. For new movies, Figures 2(a) and 2(b) show that most of the datapoints are below the dotted lines, which means users in Twitter are in general more positive towards the movies considered[5].

**Quantitative Results.** Here we introduce a set of three metrics $(\mathcal{P}, \mathcal{B}, \mathcal{I})$ to quantitatively summarize the discrepancy across two sets of positive review proportions. Let $n$ be the number of movies considered, $x_i$ be the positive proportion for the $i$-th movie in Twitter, and $y_i$ be that in IMDb or Rotten Tomatoes. The metrics $\mathcal{P} \in [0,1]$ and $\mathcal{B} \in [-1,1]$ are defined using the *median proportion* $(x^*, y^*)$, where $x^* = \text{median}\{x_1, \ldots, x_n\}$ and $y^* = \text{median}\{y_1, \ldots, y_n\}$. Then we have

$$\mathcal{P} = \frac{x^* + y^*}{2}, \quad \mathcal{B} = 1 - \tan^{-1}\left(\frac{y^*}{x^*}\right) \bigg/ \frac{\pi}{4}.$$

$\mathcal{P}$ is the **Positiveness** of the combined population of Twitter and IMDb/Rotten Tomatoes users in terms of the median $(x^*, y^*)$. $\mathcal{B}$ is the **Bias** in positiveness of Twitter users over IMDb/Rotten Tomatoes as the distance between the median and the $\pi/4$ line.

The metric $\mathcal{I}$ applies the notion of *mutual information* from information theory [3]. Let the interval $[0,1]$ be divided into $m$ subintervals: $b_1 = [0, a_1], b_2 = (a_1, a_2],$

---

[4] The only exceptions were *Haywire* and *Star Wars I: The Phantom Menace 3D*, for which $P/N < 1$ as the audience and critic responses were uniformly negative.

[5] Recall the ratio $\frac{AP}{AP+AN}$ is an underestimate of the actual proportion for Twitter, so the datapoints should be even further below the dotted lines, and our results still hold.

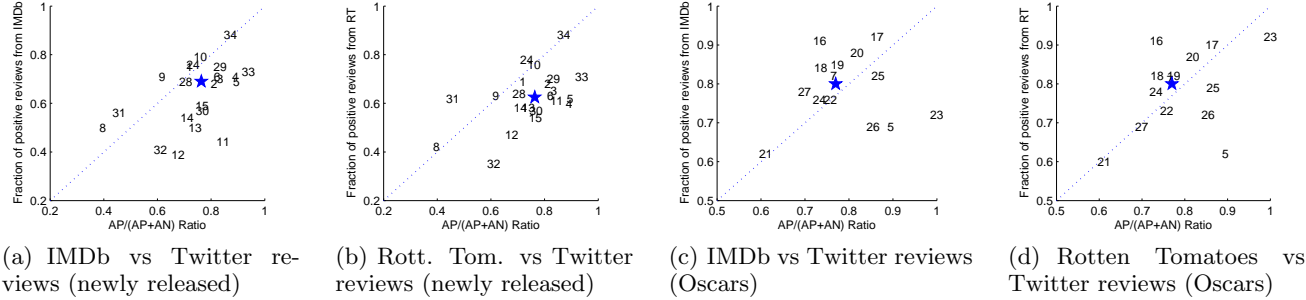

(a) IMDb vs Twitter reviews (newly released)

(b) Rott. Tom. vs Twitter reviews (newly released)

(c) IMDb vs Twitter reviews (Oscars)

(d) Rotten Tomatoes vs Twitter reviews (Oscars)

Figure 2: (a) and (b) show that the new movies score more positively from Twitter users than the general population of IMDb and Rotten Tomatoes, and (c) and (d) show that the oscar-nominated movies generally score more positively in IMDb and Rotten Tomatoes than from Twitter users

| Comparison | $\mathcal{P}$ | $\mathcal{B}$ | $\mathcal{I}$ |
|---|---|---|---|
| Twitter-RT Oscars | 0.79 | -0.024 | 0.56 |
| Twitter-IMDb Oscars | 0.79 | -0.024 | 0.42 |
| Twitter-RT Newly Released | 0.69 | 0.13 | 0.67 |
| Twitter-IMDb Newly Released | 0.73 | 0.064 | 0.52 |

Table 5: Summary metrics.

$\ldots, b_m = (a_m, 1]$. Then $\mathcal{I}$ is defined as

$$\mathcal{I} = \sum_{i=1}^{m}\sum_{j=1}^{m} p_{XY}(i,j) \log_2 \frac{p_{XY}(i,j)}{p_X(i) p_Y(j)},$$

where 
$p_X(i) = \#\{(x_k, y_k): x_k \in b_i\}/n$
$p_Y(j) = \#\{(x_k, y_k): y_k \in b_j\}/n$
$p_{XY}(i,j) = \#\{(x_k, y_k): x_k \in b_i, y_k \in b_j\}/n.$

Intuitively, $\mathcal{I}$ quantifies the **Inferrability** across different sets of reviews, i.e., if one knows the average rating for a movie in Twitter, how accurately he/she can use it to predict the average rating on IMDb. This is intrinsically related to the spread of datapoints in the scatter plots. For example, if there are many movies with $x_i$ in some small range but at the same time they have very different $y_i$, knowing a movie to have $x_i$ in that range does not help much in predicting its $y_i$ value.

We compute the three metrics for the four pairs of ratings with results shown in Table 5 ($\mathcal{I}$ is computed by dividing $[0,1]$ into ten equal-sized subintervals). The metrics capture what we can observe from the scatter plots more concisely: (1) Oscars-nominated movies have higher overall ratings (higher $\mathcal{P}$), and (2) Twitter users are more positive towards newly released movies ($\mathcal{B} > 0$). Not obvious from the plots is the ratings in Rotten Tomates are closer to Twitter by a higher $\mathcal{I}$.

### 4.3 Can Twitter Hype predict Movie Ratings?

IMDb and Rotten Tomatoes' user ratings[6] are often used as a predictors of a movie's quality and box-office potential. With the ready availability of OSN user opinion as poll data, researchers have proposed using pre-release "hype" on Twitter, measured by the number of tweets about a movie before its release, to estimate the opening day box-office [1]. We extend this notion of hype to a more generic metric of *hype-approval factor* to study how well such pre- and post- release hype on Twitter correspond to a movie's eventual ratings from the general population on IMDb and Rotten Tomatoes.

Given our ability to classify positive tweets into those that were made before watching (i.e., in hype) and after watching (i.e., in approval) a movie, we can measure their ratio as the **hype-approval factor**, $\mathcal{H}$:

$$\mathcal{H} = \frac{BP}{AP} = \frac{\text{\# Positive tweets before watching}}{\text{\# Positive tweets after watching}}$$

Using tweets collected over a period of time (e.g., a month), if the ratio of $\frac{BP}{AP} \approx 1$, then it indicates that the movie lived well up to its hype. A ratio less than 1 indicates that a movie generated much less hype than its post-release audience approval, while a ratio greater than 1 is indicative of a high hype that may be further heightened by audience approval over time.

Figure 3(a) (Figure 3(b))[7] shows the relationship between the fraction of positive ratings for different movies from IMDb and Rotten Tomatoes users versus their Twitter hype-approval factor, $\mathcal{H}$ (hype count, $BP$). From these plots, we see that for either metric, there are several movies with low $\frac{BP}{AP}$ (and low $BP$) that get very high scores in both IMDb and Rotten tomatoes (e.g., *Chronicle* (10), *The Secret World of Arrietty* (34)). On the other hand, some movies that enjoy a higher $\frac{BP}{AP}$ (and/or high $BP$) in Twitter can get lower ratings from the general population (e.g., *The Vow* (28)).

This reaffirms the observation from Figures 2(a) to 2(d) that we need to be cautious in drawing conclusions about a movie's success from observed Twitter trends. Even accounting for the hype and the approval level in Twitter may be insufficient to predict a movie's rating from the general online population.

---

[6]For a fair comparison, we exclude scores from movie critics.

[7]In order to have sufficient datapoints across all newly released movies, for these two figures and Figure 4, we track tweets from the week after release.

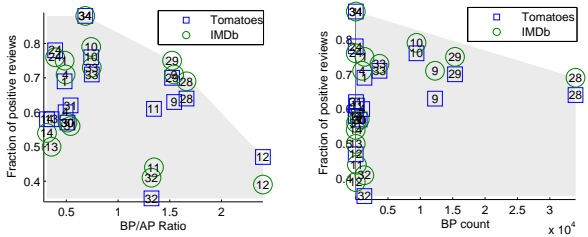

(a) Online Ratings vs. $\mathcal{H}$  (b) Online Ratings vs. BP

**Figure 3: Fraction of positive ratings from IMDb and Rotten Tomato versus (a) hype-approval factor, $\mathcal{H}$, and (b) hype, $BP$, in Twitter**

### 4.4 Box-office Gains: Twitter Hype-satisfaction or IMDb Ratings?

Earlier works [1] have reported that a higher number of positive tweets or "hype" about a movie on Twitter directly translates into a higher box-office sales on the opening day. However, whether a good box-office sale is sustained or not also depends on the amount of positive tweets made by satisfied Twitterers after watching the movie (i.e., AP tweets), which in turn can induce more hype (i.e., BP tweets). We show in Figure 3(a) that a high (low) $\frac{BP}{AP}$ ratio does not necessarily correspond to a high (low) rating for a movie in the other sites, and hence, it is of interest to explore whether such scores are any good indicators of a movie's eventual box-office.

Figure 4 shows the classification of the movies listed in Table 1 by their Twitter's $\frac{BP}{AP}$ ratio in the first level, by their IMDb scores in the second level, and finally by their box-office figures from IMDb. A box-office earning of $50 million is taken as a standard valuation for financial success, although the key observations reported below will hold for any amount between $20 million to $60 million box-office for the given list of movies. The figure highlights a few interesting outcomes:

(a) Even if a movie has $\frac{BP}{AP} < 1$ (low hype-approval) and *IMDb rating < 0.7* (low-score), it can still become financially successful (e.g., *Journey 2* (30)).

(b) Movies that have $\frac{BP}{AP} < 1$ (low hype-approval) but *IMDb rating > 0.7* (high-score), or $\frac{BP}{AP} > 1$ (high hype-approval) but *IMDb rating < 0.7* (low-score), can be financially either successful or unsuccessful.

(c) None of the movies with $\frac{BP}{AP} > 1$ and *IMDb rating > 0.7* have a box-office success of less than $50M.

In other words, a high score on IMDb, complemented with a high hype-approval factor in Twitter, can be indicative of financial success, but otherwise marketers need to be careful about drawing conclusions regarding the net box-office outcome for a movie.

### 5. CONCLUSIONS

This paper presents a study that compares data from Twitter to other online user population. We show that

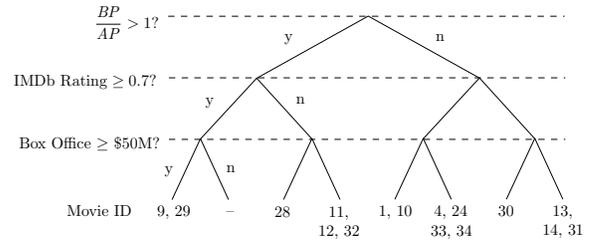

**Figure 4: Relationship between IMDb ratings, Twitter review scores, and Box-office outcomes**

Twitter users are more positive in their reviews across most movies in comparison to other rating sites, like IMDb and Rotten Tomatoes. Moreover, compared to IMDb and Rotten Tomatoes users, the computed scores from Twitter users are slightly less positive for the Oscar-nominated best films but more positive for non-nominated films, which we quantify by introducing a few interesting information-theory inspired metrics $\mathcal{P}$, $\mathcal{B}$, and $\mathcal{I}$, which together capture the "bias" observed among Twitter users. We also introduce a hype-approval metric $\mathcal{H}$, measured as a ratio of the total number of positive tweets the users make before and after watching a movie, and relate it with the ratings for the movie on IMDb or Rotten Tomatoes. Finally, we show that scores computed from Twitter reviews and other online sites do not necessarily translate into predictable box-office.